# Impurity states of electrons in quantum dots in external magnetic fields

A.M. Ermolaev, G.I. Rashba

Department of Theoretical Physics, V.N. Karazin Kharkiv National University,

4 Svobody square, 61077, Kharkiv, Ukraine

Email: georgiy.i.rashba@univer.kharkov.ua

The influence of isolated impurity atoms on the electron energy spectrum in a parabolic quantum dot in quantizing magnetic field is studied. The impurity potential is approximated by a Gaussian separable operator which allows one to obtain the exact solution of the problem. We demonstrate that in the electron energy spectrum there is a set of local levels which are split from the Landau zone boundaries in the upward or downward direction depending on the impurity type. We have calculated the local level positions, the wave functions of electrons in bound states, and the residues of the electron scattering amplitudes by impurity atoms at the poles.

**PACS:** 73.21.La, 71.70.Di, 71.55.-i
**Keywords:** impurity atoms, quantum dots, impurity states, local levels, residues of the electron scattering amplitudes

Physicists and technologists are interested in quantum dots [1] for a number of reasons, the main one of which is that they are functional elements of modern devices and electronic gadgets. This interest has intensified since experimentalists learnt how to synthesize these nanosystems in laboratories. For theorists, quantum dots are of interest because they are convenient objects to test new calculation methods. Of particular interest are properties of the quantum dots with impurity atoms in a magnetic field. Due to a small number of electrons in the quantum dot even a single impurity atom influences strongly its properties. Within a limited volume of the quantum dot in a magnetic field, interesting phenomena become present like hybridization of spatial and magnetic quantization of the electron's motion. The electron localization on individual impurity atoms considered here is one of such phenomena.

The impurities in bulk- and nanosystems play a double role. On the one hand, they determine the low-temperature behaviour of the kinetic characteristics of a system. On the other hand, the impurities influence the energy spectrum of the system thereby leading to the appearance of local and resonant levels in the energy spectrum of quasi-particles [2-4]. These levels exert a profound influence on the properties of solids [2-9].

In this work, we present the results of a theoretical study made on the local states of electrons in quantum dots in a quantizing magnetic field using the method of degenerate regular perturbations [2]. We shall use the model of a parabolic quantum dot [1] based on 2D electron gas with the confinement potential $m_* \omega_0^2 \rho^2 / 2$, where $m_*$ is the effective electron mass, $\omega_0$ is the potential parameter, $\rho$ is the polar radius in the plane $z = 0$ occupied by the

2D electron gas. The magnetic field $B$ with the vector potential $\vec{A} = [\vec{B}\vec{\rho}]/2$ is perpendicular to this plane.

The wave function of the electron's stationary state in the field of confinement and in the magnetic field is equal to [10-12]:

$$\psi_{nm}(\rho,\varphi) = \frac{1}{l}\sqrt{\frac{n!}{(n+|m|)!}}\left(\frac{\rho^2}{2l^2}\right)^{|m|/2}\exp\left(-\frac{\rho^2}{4l^2}\right)L_n^{|m|}\left(\frac{\rho^2}{2l^2}\right)\frac{e^{im\varphi}}{\sqrt{2\pi}}, \qquad (1)$$

where $n$ and $m$ are the radial and azimuthal quantum numbers, $\varphi$ is the polar angle, $l = (m_*\omega)^{-1/2}$ is the magnetic length, $\omega = (\omega_c^2 + 4\omega_0^2)^{1/2}$ is the hybrid electron frequency ($\omega_c$ denotes the cyclotron frequency), $L_n^{|m|}$ are the generalized Laguerre polynomials. The quantum constant here and below is taken to be equal to unity. The electron energy in the state (1) is:

$$\varepsilon_{nm\sigma} = \omega\left[n + \frac{1}{2} + \frac{1}{2}\left(|m| - \frac{\omega_c}{\omega}m\right)\right] + \sigma\mu B, \qquad (2)$$

where $\sigma = \pm 1$ denotes the spin quantum number, $\mu$ is the electron spin magnetic moment. The spectrum (2) is of serial nature [1].

Let us assume that the impurity atom is located at the center of the coordinate system. The impurity potential is approximated by separable operator

$$u = |\eta\rangle u_0 \langle\eta|, \qquad (3)$$

where $u_0$ is the constant, while $|\eta\rangle$ is a certain normalized state. Let us consider the function $\langle\vec{\rho}|\eta\rangle$ to have a Gaussian form:

$$\langle\vec{\rho}|\eta\rangle = (\sqrt{\pi}a)^{-1}\exp\left(-\frac{\rho^2}{2a^2}\right), \qquad (4)$$

where $a$ is the extent of this function. Fig.1 shows the potential $V$ of confinement and impurity potential of attraction (a) and repulsion (b) of electron in the quantum dot. The potential (3) was used in the references [2-4], the function (4) was used in studying the electron impurity states in bulk conductors in the reference [13]. The limit passage to the $\delta$-potential $\upsilon_0 \delta(\vec{\rho})$ is made by the substitution

$$4\pi \lim_{\substack{a \to 0 \\ u_0 \to \infty}} \left(a^2 u_0\right) = \upsilon_0. \quad (5)$$

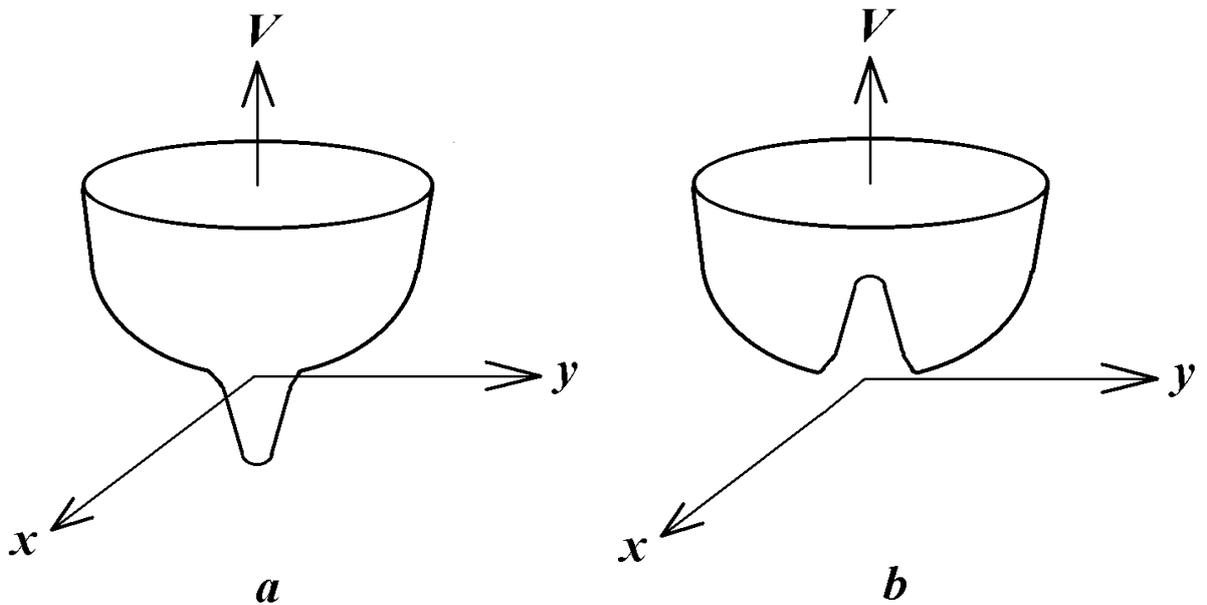

Fig.1. Potential of confinement and impurity atom potential of attraction (a) and repulsion (b) of electron in the quantum dot.

The advantages of the chosen potential (3), (4) are such that it allows for the exact solution of the problem of the electron impurity states in the quantum dots and it helps to avoid divergences of sums and integrals which are present in the case of dot potential. Besides, in the theory there are two independent parameters $u_0/\omega$ and $a/l$, but not just one, $\upsilon_0$, as it is in the case of the zero radius potential method [14].

The Green function $G$ of electron in the quantum dot is associated with the scattering operator $T$ via the relationship [3] $G = G_0 + G_0 T G_0$, where $G_0$ is the Green function in the absence of the impurity potential. The operator $T$ is equal to [15] $|\eta\rangle T_\sigma(\varepsilon)\langle\eta|$, where the function

$$T_\sigma(\varepsilon) = u_0\left(1 - u_0 \sum_\kappa \frac{|\langle\eta|\kappa\rangle|^2}{\varepsilon - \varepsilon_{\kappa\sigma} + i0}\right)^{-1} = u_0\left\{1 - u_0\left[D_\sigma(\varepsilon) - i\pi g_\sigma(\varepsilon)\right]\right\}^{-1} \quad (6)$$

is proportional to the scattering amplitude of electron with the energy $\varepsilon$ by the impurity atom. Here, $\kappa = (n,m)$, $\varepsilon_{\kappa\sigma}$ are the electron energy levels (2). The overlap of states $\langle\eta|\kappa\rangle$ entering equation (6) is given by

$$\langle\eta|nm\rangle = \delta_{m0}\sqrt{2}\frac{l}{a}\left(\frac{l^2}{a^2}+\frac{1}{2}\right)^{-1}\left[\left(\frac{l^2}{a^2}-\frac{1}{2}\right)\left(\frac{l^2}{a^2}+\frac{1}{2}\right)^{-1}\right]^n. \quad (7)$$

The function $g_\sigma(\varepsilon)$ has the following form:

$$g_\sigma(\varepsilon) = \frac{8}{\omega}\left(\frac{l}{a}\right)^2\left[2\left(\frac{l}{a}\right)^2+1\right]^{-2}\left[\left(2\frac{l^2}{a^2}+1\right)\left(2\frac{l^2}{a^2}-1\right)\right]^{2\upsilon_\sigma(\varepsilon)-1}\sum_{n=0}^{\infty}\delta[n+\upsilon_\sigma(\varepsilon)], \quad (8)$$

where

$$\upsilon_\sigma(\varepsilon) = \frac{1}{2} + \frac{\sigma\mu B - \varepsilon}{\omega}. \quad (9)$$

In the limiting case $u_0 \to \infty$, $a \to 0$ the factor $u_0 g_\sigma(\varepsilon)$ is equal to $\upsilon_0 \nu_\sigma(\varepsilon)$, where $\nu_\sigma(\varepsilon)$ is the density of electron states in magnetic field. The function $D_\sigma(\varepsilon)$ is related to the function (8) via the relationship:

$$D_\sigma(\varepsilon) = \text{P.} \int_{-\infty}^{\infty} d\varepsilon' \frac{g_\sigma(\varepsilon')}{\varepsilon - \varepsilon'}.$$

The Kronecker symbol $\delta_{m0}$ in (7) indicates that the impurity potential (3), (4) scatters the electrons with zero projection of the angular moment on the magnetic field direction only.

The scattering amplitude poles (6) in the complex energy plane are associated with local and resonant levels of the electron energy in the impurity atom field. They are the roots of the Lifshitz equation [3]:

$$1 - u_0 D_\sigma(\varepsilon) = 0. \tag{10}$$

In the considered case this equation is as follows:

$$\frac{\omega}{u_0} = (z-1)\Phi(z,1,\upsilon_\sigma), \tag{11}$$

where

$$z = \left(\frac{2\xi - 1}{2\xi + 1}\right)^2, \qquad \xi = \left(\frac{l}{a}\right)^2,$$

$$\Phi(z,1,\upsilon) = \sum_{n=0}^{\infty} \frac{z^n}{n+\upsilon} \tag{12}$$

$(z<1, \upsilon \neq 0, -1, -2, ...)$. The function $\Phi$ is related to the hyper-geometrical Gaussian function F via the relationship [16]

$$\Phi(z,1,\upsilon) = \frac{1}{\upsilon} F(1, \upsilon; 1+\upsilon; z). \tag{13}$$

The exact equation (11) for the local levels $\varepsilon_l$ contains two independent parameters: $\omega/u_0$ and $l/a$. The analysis of this equation indicates that in the electron energy spectrum there is a set of local levels, $\varepsilon_l$, detached from the boundaries $\varepsilon_{n0\sigma}$ of the series (2) in the downward direction in the case of attractive impurity potential $(u_0 < 0)$ and in the upward direction in the case of repulsive one $(u_0 > 0)$. Let us consider solutions of equation (11) in several limiting cases.

If $z \leq 1$, i.e. the values $a$ and $l$ differ considerably, then one can make use of the known representation of the hyper-geometric Gaussian function [16]

$$F(a,b;a+b;z) = \frac{\Gamma(a+b)}{\Gamma(a)\Gamma(b)} \sum_{n=0}^{\infty} \frac{\Gamma(a+n)\Gamma(b+n)}{\Gamma(a)\Gamma(b)(n!)^2} \times$$

$$\times \left[2\psi(n+1) - \psi(a+n) - \psi(b+n) - \ln(1-z)\right](1-z)^n$$

$$(|\arg(1-z)| < \pi, |1-z| < 1),$$

where $\psi$ denotes the logarithmic derivative of the gamma-function $\Gamma$. By using this representation, from (11) and (13) at $z \leq 1$ we obtain an approximate equation for the local levels:

$$\psi(\upsilon_\sigma) = \ln\left\{e^{-\gamma} \exp\left[\frac{\omega}{u_0(1-z)}\right](1-z)^{-1}\right\}, \tag{14}$$

where $\gamma = 0,577...$ is the Euler number. To obtain the solution of this equation at $a \ll l$ and $a \gg l$, we shall use the representation of the function $\psi$ as [16]:

$$\psi(\upsilon) = -\gamma + (\upsilon - 1) \sum_{n=0}^{\infty} \frac{1}{(n+1)(n+\upsilon)}.$$

Then from Eq.(14) we shall obtain the expression for the separation $\Delta_{n\sigma} = \varepsilon_{n0\sigma} - \varepsilon_{n\sigma}^l$ between the Landau zone boundary $\varepsilon_{n0\sigma}$ (2) and the local level $\varepsilon_{n\sigma}^l$:

$$\Delta = \begin{cases} \left\{-\omega\left\{\ln\left[\frac{1}{2}\left(\frac{l}{a}\right)^2\right] + \frac{1}{2}\left(\frac{l}{a}\right)^2 \frac{\omega}{u_0}\right\}\right\}^{-1}, & l \gg a, \\ \left\{-\omega\left\{\ln\left[\frac{1}{8}\left(\frac{a}{l}\right)^2\right] + \frac{1}{8}\left(\frac{a}{l}\right)^2 \frac{\omega}{u_0}\right\}\right\}^{-1}, & l \ll a. \end{cases} \quad (15)$$

Hence it follows that the impurity-detached local levels exist in the absence of magnetic field as well. The latter enhances the confinement potential. In a weak magnetic field the value of $\Delta$ increases with increasing value of $B$ in proportion to $\omega$, while in a strong field it decreases in accordance with the law $\Delta \sim \omega^{-1}$. When the confinement potential is absent $(\omega_0 = 0, \omega = \omega_c)$, the formulae (15) describes the positions of the local levels in the 2D electron gas in magnetic field [17,18]. It follows from the formula (15) that at $u_0 > 0$ the local level that is detached from the lower limit of the spectrum $\varepsilon_{00(-1)}$ is positioned in the region $\varepsilon < 0$, if $\mu B > \omega/2$ and $u_0 < D_{-1}^{-1}(0)$. Note that the first formula (15) for the 2D electron gas in a weak magnetic field is different by numerical factors only from the one obtained in reference [19] using another impurity potential model. The formulae (15) are readily derivable from the equation (11), if we extract in the function $\Phi$ the summand $-\ln(1-z)$ [16], leaving only the term with the minimum denominator in the remaining sum at $|\Delta| \ll \omega$. Then

$$\Delta_n = -\omega z^n \left[\ln(1-z)^{-1} + \frac{\omega}{u_0(1-z)}\right]. \quad (16)$$

If the superscript $n$ of the local level is increasing, it approaches the boundary of the series $\varepsilon_{n0\sigma}$. In the quantum limit of the equation (11) we obtain:

$$\Delta = -2u_0 \left(\frac{l}{a}\right)^2 \left[\frac{1}{2} + \left(\frac{l}{a}\right)^2\right]^{-2}.$$

If $|u_0| \to \infty$, the equation (11) comes out with $\varepsilon_l = -|u_0|$.

In the case of weak quantization of the levels $\omega \ll |\varepsilon|$ the sum in (11) can be replaced by the integral. Then the equation (11) takes the following form:

$$\frac{\omega}{u_0(1-z)} = \exp(\upsilon|\ln z|)\operatorname{Ei}(-\upsilon|\ln z|), \quad (17)$$

where $\operatorname{Ei}(x)$ – the integral exponential function [16]. At $|\upsilon \ln z| \ll 1$ one can restrain himself to the main terms of expansion of the function $\operatorname{Ei}(x)$ over the powers of $x$. Then the equation (17) takes the following form:

$$\frac{\omega}{u_0(1-z)} = \ln(|\upsilon \ln z|).$$

Hence we obtain

$$|\upsilon| = |\ln z|^{-1} \exp\left[\frac{\omega}{u_0(1-z)}\right].$$

If $u_0 < 0$ and $\omega_0 = \omega_c = 0$, then from this formula we derive the position of the local level in the 2D electron gas

$$\varepsilon_l = -\frac{1}{2m_*a^2}\exp\left(-\frac{1}{2m_*a^2|u_0|}\right).$$

In the limit of the $\delta$-potential (5) the exponent therein coincides with the one obtained in reference [20].

It follows from the equation (11) that in the $\varepsilon_l$ vs. magnetic field chart there is a sheer dip at $\xi = 1/2$ in the field

$$B_0 = \frac{2m_*c}{e}\left(\frac{1}{m_*^2 a^4} - \omega_0^2\right)^{1/2}. \qquad (18)$$

Here $e$ is the electron charge value, $c$ is the speed of light. In order to find the local level position at $\xi \approx 1/2$, we shall use the asymptotic of the integral exponential function

$$\text{Ei}(-x) \sim -\frac{e^{-x}}{x}$$

where $|\upsilon \ln z| \gg 1$. Then we find from the formula (17) at $u_0 < 0$ that

$$\Delta = |u_0|\left|\ln\left(\frac{1}{2}-\xi\right)^2\right|^{-1}. \qquad (19)$$

Fig.2 shows the dependence of the relative separation $\Delta/\omega$ between the levels $\varepsilon_{n0\sigma}$ (2) and the local levels on magnetic field at $|u_0|/\varepsilon_0 = 0.25$, $\omega_0/\varepsilon_0 = 0.25$. Here $\varepsilon_0 = (m_*a^2)^{-1}$, $x = \omega_c/2\omega_0$, $x_0 = eB_0/2m_*c\omega_0$. The solid curves are obtained from the asymptotes (15) and (19), the dashed lines show schematically the dependence in the non asymptotic regimes.

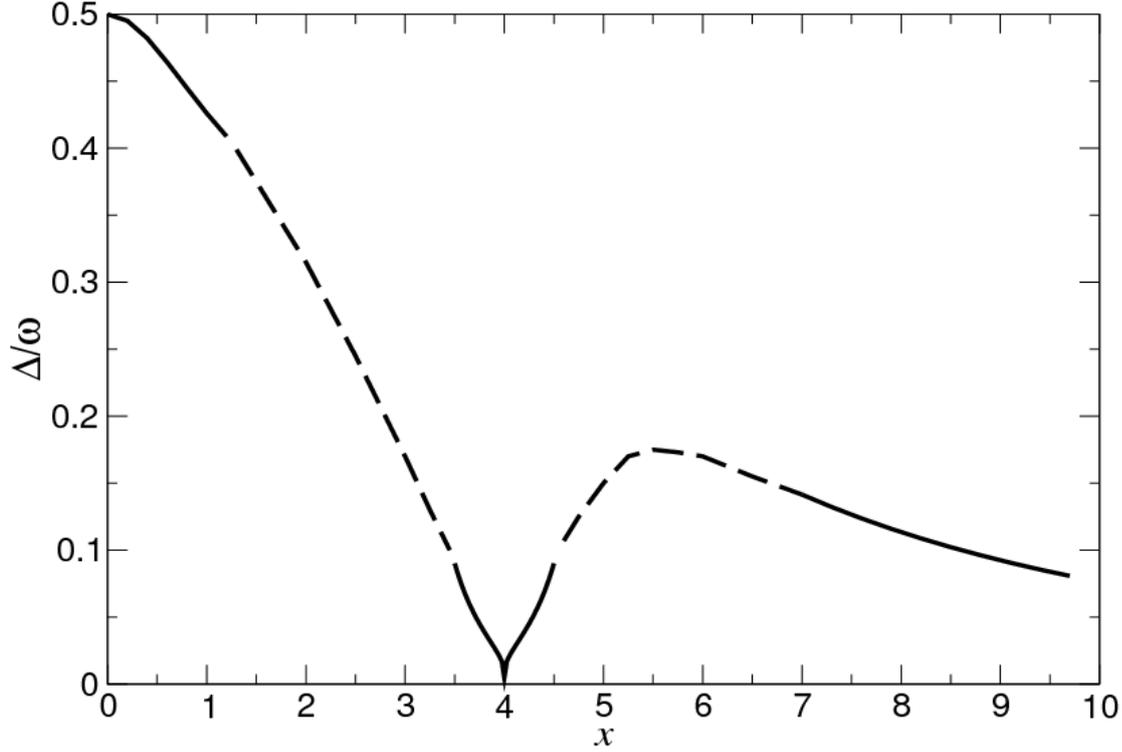

Fig.2. Dependence of the separation between levels (2) and local levels on the magnetic field.

The normalized electron radial wave function in the bound state is

$$R_\sigma(\rho) = \frac{1}{l} \frac{|\Gamma(\upsilon_\sigma)|}{[\psi'(\upsilon_\sigma)]^{1/2}} \exp\left(-\frac{\rho^2}{4l^2}\right) \Psi\left(\upsilon_\sigma, 1; \frac{\rho^2}{2l^2}\right),$$

where $\psi'$ is the three-gamma function, $\Psi$ is the degenerate hyper-geometric function [16]. The value $\upsilon_\sigma$ in this formula is taken in the point $\varepsilon = \varepsilon_l$.

In the applications we shall have to deal with residues of the scattering amplitude (6) at the poles. They are as follows:

$$r = \begin{cases} \dfrac{1}{2}\omega^2 \left(\dfrac{l}{a}\right)^2 \zeta^{-1}\left(2, -\dfrac{\Delta}{\omega}\right), & a \ll l, \\ \dfrac{1}{8}\omega^2 \left(\dfrac{a}{l}\right)^2 \zeta^{-1}\left(2, -\dfrac{\Delta}{\omega}\right), & a \gg l, \end{cases}$$

where $\zeta$ is the generalized Riemann dzeta-function. As in the bulk conductors [8], the residues describe, in particular, the forces of oscillators of the resonant optical transitions of electrons between levels (2) and local levels.

For numeric evaluations we use a value, which is typical for the quantum dot based on the heterosystem $GaAs/Al_xGa_{1-x}As$ with the 2D electron gas, of $m_* = 10^{-28}$ g. Then for $a/l = 0,1$, $\omega_0 = \omega_c$, $|u_0|/\omega = 10$ in the field of $B = 10^4$ G from (15) we obtain $|\Delta|/\omega = 0,1$.

**Conclusion**

Since the number of electrons in a quantum dot is small, its properties are sensitive to external influence. In a quantizing magnetic field the spectrum of the Landau electrons in the quantum dot becomes serial. Even one impurity atom may have a considerable effect on the spectrum. We have demonstrated that the impurity atoms of donor and acceptor types split off the local levels from the Landau levels in the quantum dot. The split-off goes down and up depending on the type of impurity. To calculate the value of the split-off $\Delta$, we used the model of the parabolic quantum dot and of the Gaussian separable impurity potential that allows for the exact solution of the problem. We found the positions of the local levels in the regimes of weak $(a \ll l)$ and strong $(a \gg l)$ magnetic fields. The minimum of the split-off is found in the region of intermediate strengths of the field. As in the bulk conductors [8], the local levels influence the components of tensor of high-frequency permittivity of electrons, and they can manifest themselves in experiments on the electromagnetic radiation absorption in quantum dots.

This work was partially supported by the INTAS program (grant INTAS-01-0791). We would like to thank T. Rashba for help with preparation of the manuscript.